\documentclass[11pt]{article} 

\usepackage{authblk}  
\usepackage[utf8]{inputenc}      
\usepackage[T1]{fontenc}         
\usepackage{xcolor}
\usepackage{amsmath, amssymb, amsthm} 
\usepackage{graphicx}            
\usepackage{hyperref}            
\usepackage{pgfplots}
\pgfplotsset{compat=1.18}
\usepackage{siunitx} 
\usepackage{float}
\usepackage{geometry}            
\usepackage{booktabs}
\usepackage{booktabs}   
\usepackage{multirow} 
\usepackage{subcaption} 

\title{\textbf{Towards Enterprise-Ready Computer Using Generalist Agent}}

\author{\textbf{Sami Marreed}$^{*}$, \textbf{Alon Oved}$^{*}$, \textbf{Avi Yaeli}$^{*}$, \textbf{Segev Shlomov}$^{*}$, \textbf{Ido Levy}$^{*}$, \textbf{Offer Akrabi},  \textbf{Aviad Sela}, \textbf{Asaf Adi}, \textbf{Nir Mashkif}}

\affil{\small IBM Research}

\affil{\small \texttt{\{sami.marreed, alon.oved, segev.shlomov1, ido.levy1\}@ibm.com}}

\affil{\small \texttt{\{aviy, offer.akrabi, sela, adi, nirm\}@il.ibm.com}} 

\date{}

\begin{document}

\maketitle

\begingroup
\renewcommand\thefootnote{}\footnotetext{*These lead authors  contributed equally to this work}
\endgroup

\begin{abstract}
This paper presents our ongoing work toward developing an enterprise-ready Computer Using Generalist Agent (CUGA) system. Our research highlights the evolutionary nature of building agentic systems suitable for enterprise environments. By integrating state-of-the-art agentic AI techniques with a systematic approach to iterative evaluation, analysis, and refinement, we have achieved rapid and cost-effective performance gains, notably reaching a new state-of-the-art performance on the WebArena and AppWorld benchmarks. We detail our development roadmap, the methodology and tools that facilitated rapid learning from failures and continuous system refinement, and discuss key lessons learned and future challenges for enterprise adoption.

\end{abstract}

\section{Introduction}

The development of enterprise-ready, computer-using generalist agents represents a significant frontier in artificial intelligence, poised to revolutionize productivity, workflows, automation, and decision-making across diverse industries. Recent advances in large language, vision, and action models, coupled with progress in agentic AI frameworks and implementations, are continuously raising the bar on existing computer-using benchmarks. While recent announcements, such as Anthropic's Computer Use \cite{anthropicComputerUse2024} and OpenAI's Operator \cite{openaiOperator2025}, suggest a growing commercial opportunity, realizing this vision requires more than just cutting-edge models, algorithms, or product prototypes, and significant challenges remain.

At IBM Research, our ambition is to pioneer the development of agent systems that transcend mere task completion and encompass the full spectrum of dimensions required for enterprise adoption, such as privacy, safety, trustworthiness, and cost-effectiveness of AI agentic solutions. As part of this mission, we have begun to develop a Computer Using Generalist Agent (CUGA). Our vision for IBM CUGA is to develop a generalist agent that can be adapted and configured by knowledge workers to perform routine or complex aspects of their work in a safe and trustworthy manner. Our first version focuses on knowledge worker tasks within web applications, and we tested it on the WebArena benchmark \cite{zhou2023webarena}.

\vspace*{0.5cm}

\begin{table}[h!]
\centering
\caption{CUGA Results on Benchmarks}
\vspace*{0.3cm}
\begin{tabular}{llcl}
\toprule
\textbf{Benchmark Name} & \textbf{Success/Completion Rate} & \textbf{Domain} & \textbf{Description} \\
\midrule
WebArena             & 61.7\% success rate              & Web             & Evaluates web-based tasks \\
AppWorld             & 48.2\% scenario completion rate    & API             & Evaluates API tasks \\
\bottomrule
\end{tabular}
\label{tab:benchmark-results}
\end{table}

As shown in Table~\ref{tab:benchmark-results}, CUGA achieves a new state-of-the-art result of \textbf{61.7\%} task completion on the WebArena benchmark \cite{wearena_leaderboard}. On the AppWorld benchmark \cite{appworld}, which challenges agents to complete complex, multi-step workflows across diverse API-driven applications, the agent achieves a \textbf{46\%} scenario completion rate—also setting a new state-of-the-art. Notably, AppWorld evaluates an agent’s ability to dynamically select appropriate APIs, manage variables, reason about preconditions and outputs, and align its strategy with long-horizon goals—skills that are essential for real-world enterprise systems.

This paper outlines the current state of our work, detailing the evolution of our agentic architecture in response to the challenges posed by the WebArena and AppWorld benchmarks. We describe our iterative development methodology and the tools that facilitated rapid learning from failures and cost-effective architectural improvements. Furthermore, we share lessons learned and highlight key challenges in realizing the full potential of such systems. Our primary contribution lies in disseminating the methodology, architecture, and practical experience that enabled us to achieve top performance on both the WebArena and AppWorld leaderboards. Additionally, we have created a fully interactive dashboard\footnote{IBM CUGA dashboard: https://cuga.dev/} that showcases our performance results and agent trajectories.

\section{Methodology and Tools}

Our approach to developing an enterprise-ready computer using generalist agent (CUGA) is grounded in a philosophy of iterative evolution and rapid learning. We began with a simple agent architecture, intentionally designed to be a starting point, and committed to refining it based on empirical results and failure analysis. This evolutionary strategy enables us to adapt quickly to the complexities of real-world scenarios and continually improve our performance.

\begin{figure}[h!]
    \centering
    \includegraphics[width=1.0\textwidth]{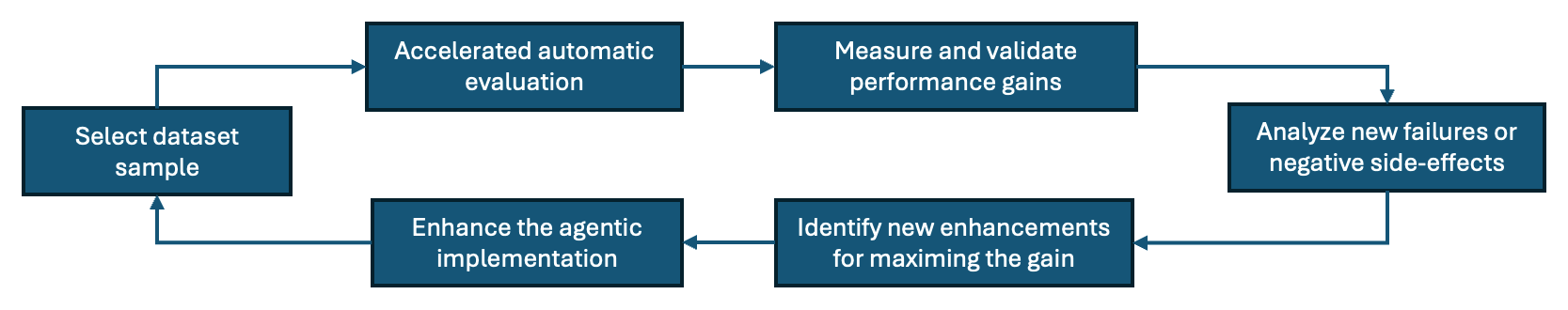}
    \caption{CUGA’s iterative evaluate-analyze-enhance cycle, enabling rapid failure diagnosis, targeted improvements, and continuous performance gains.}  
    \label{fig:cuga_lifecycle}
\end{figure}

A cornerstone of our development methodology is the implementation of a smart sampling strategy. Recognizing the time-intensive nature of evaluating agent systems on comprehensive benchmarks, we adopted a technique of selecting an initially small, representative subset of the benchmark and then enlarging the subset as the system evolved to become better and more stable. This allowed us to rapidly test hypotheses, identify failure areas and side effects, and iterate on improvements before scaling up to larger portions of the benchmark. This approach significantly accelerates our learning cycle, enabling us to achieve rapid performance gains. Figure \ref{fig:cuga_lifecycle} depicts the main phases in our iterative evaluate-analyze-enhance process. In each iteration, we evaluate the system on a larger sample, validate that expected performance gains are achieved, analyze failures, and prioritize areas of improvement that would maximize the performance gain in the following cycle.

To facilitate this iterative process, we developed a suite of evaluation and analysis tools designed to provide comprehensive insights and accelerate development:

\begin{figure}[ht!]
    \centering
    \includegraphics[width=\textwidth]{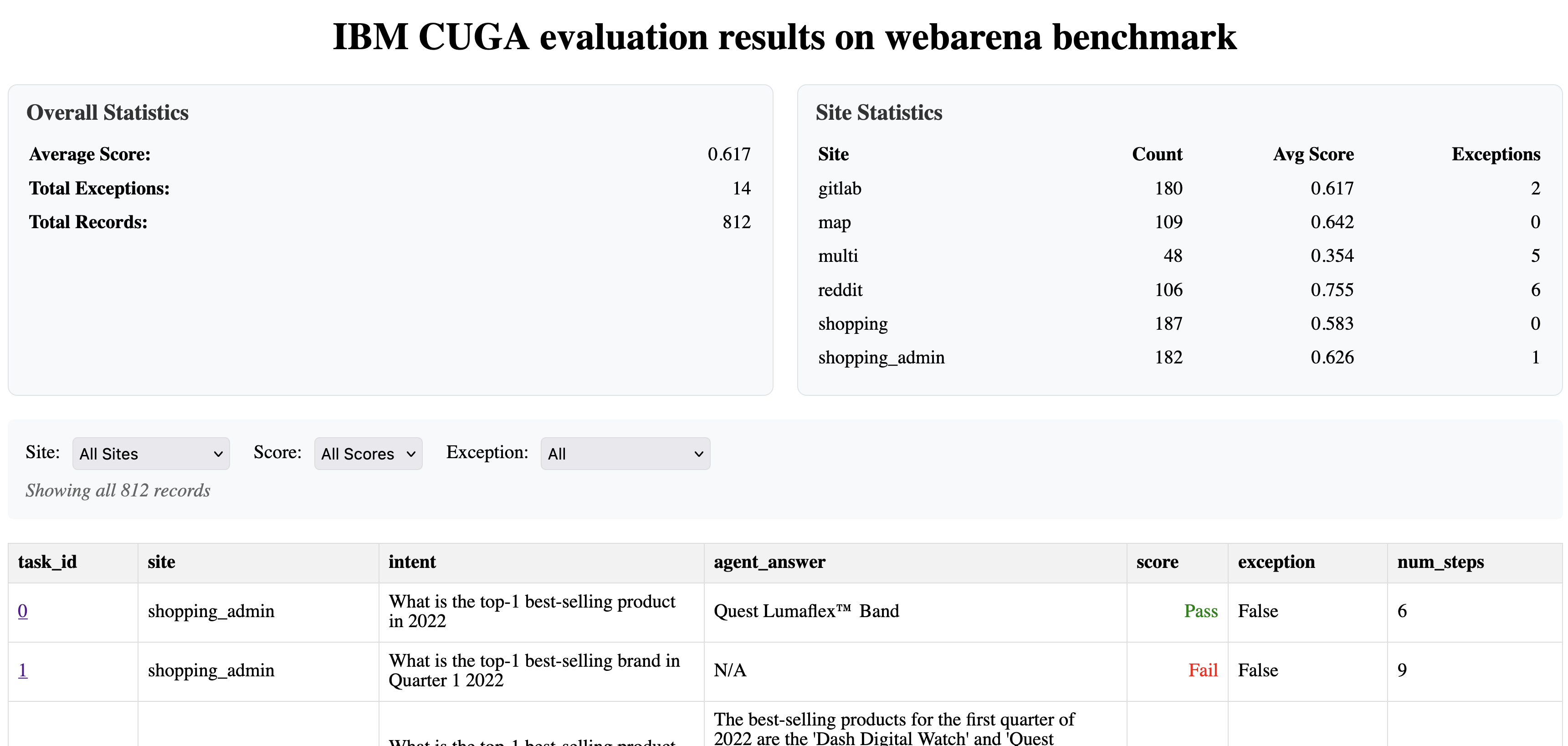}
    \caption{CUGA performance dashboard providing an overview and detailed performance results per task, with drill-down links into trajectories}
    \label{fig:cuga_dashboard} 
\end{figure}

\begin{enumerate}
    \item \textbf{Performance Dashboard:} This dashboard, as depicted by Figure \ref{fig:cuga_dashboard} provides a real-time overview of the agent's performance across various metrics. It allows us to quickly assess the impact of new versions and identify areas for improvement.
    \item \textbf{Comparative Analysis:} Building upon the performance dashboard, this view enables direct comparison of results between different agent versions. It highlights previously resolved failures that are now successful, new failures on new data points, as well as failures on previously successful runs. This feature enables the rapid assessment of the impact of changes, validation of hypotheses, detection of side effects, and facilitates regression analysis.  
    \item \textbf{Trajectory Visualization and error classification:} This tool, depicted in Figure \ref{fig:cuga_trajectory} allows us to delve into individual failure cases, visualizing the agent's interaction with the environment, its perception, reasoning process, and actions taken by different components. This detailed view enables us to quickly pinpoint and classify the root cause of failures, generating targeted hypotheses for improvement.
    \item \textbf{Parallel Execution Framework:} To significantly reduce evaluation time, we implemented a parallel execution framework. This framework allows us to run multiple evaluations concurrently, reducing evaluation times from days to hours, and from hours to minutes. This speedup is crucial for rapid iteration and experimentation.
\end{enumerate}

\begin{figure}[htbp]
    \centering

    \includegraphics[width=\textwidth]{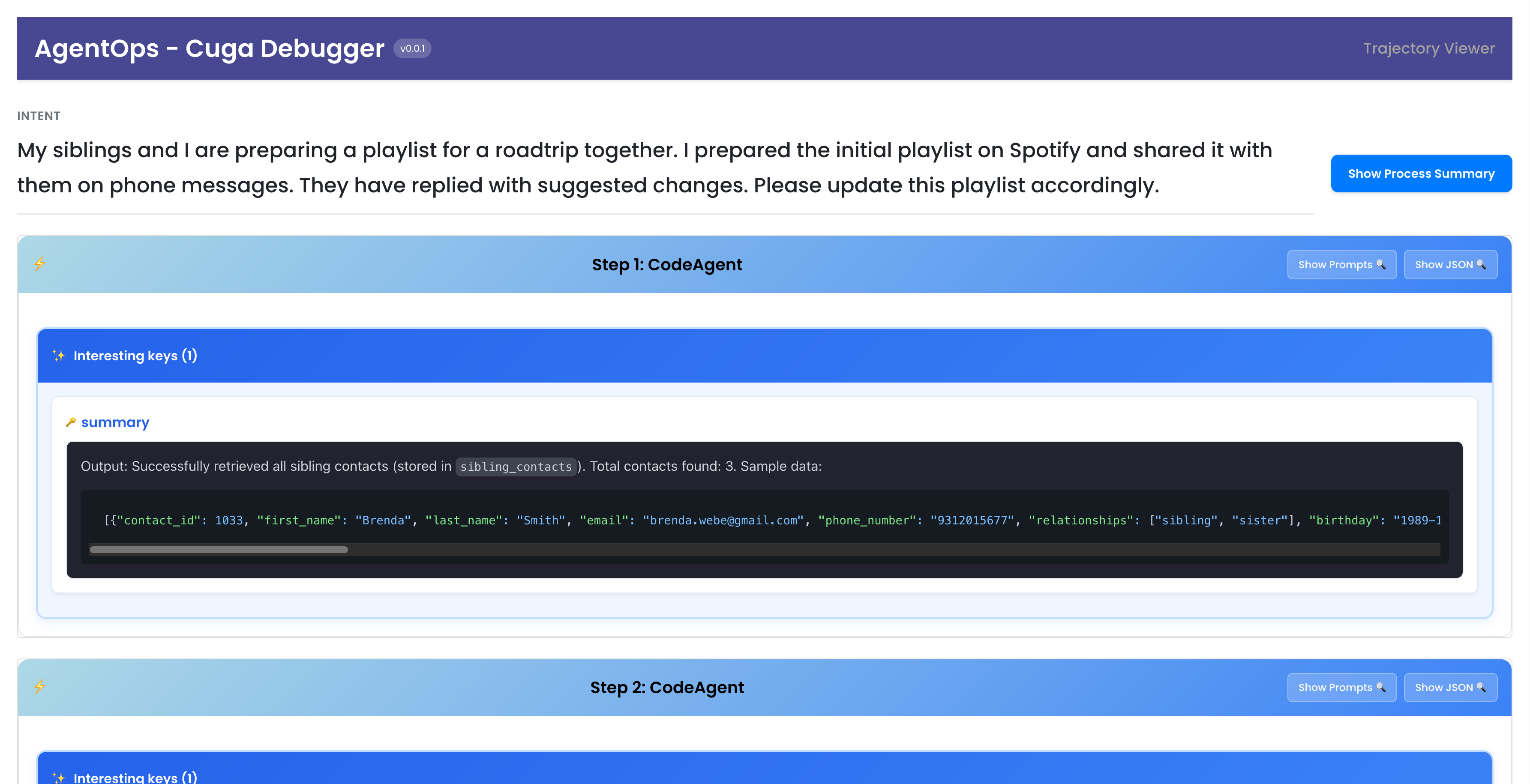}
    \caption{The CUGA trajectory visualization on the AppWorld task.}
    \label{fig:tapi_sub_agent}

    \vspace{2em} 

    \includegraphics[width=\textwidth]{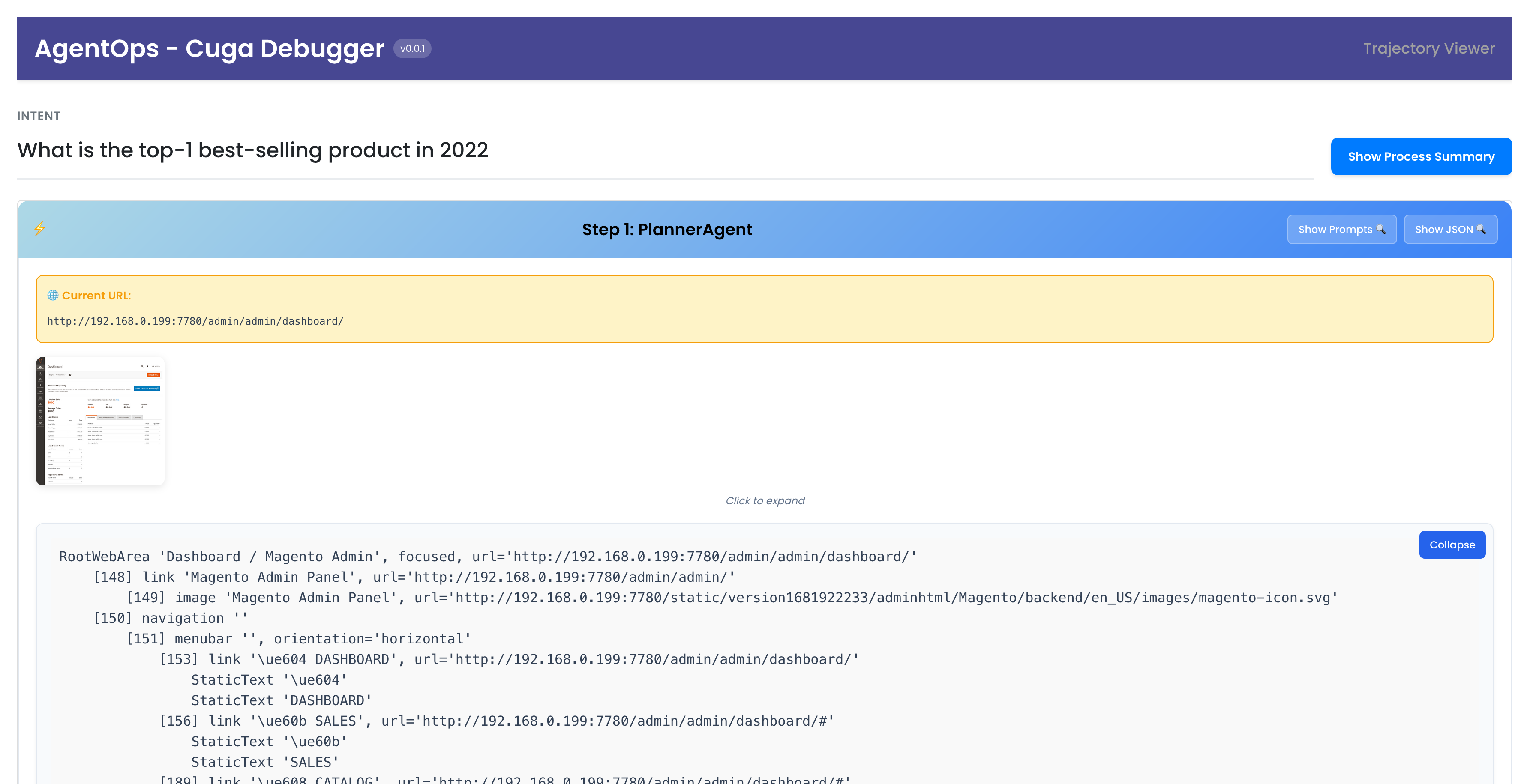}
    \caption{The CUGA trajectory visualization on the WebArena task provides easy access to the environment observation as a screenshot, perception as an accessibility tree, the action instruction, and its element grounding in the accessibility tree.}
    \label{fig:tbrowser_sub_agent}

    \vspace{2em} 

    \caption*{\textbf{Figures \ref{fig:tapi_sub_agent} and \ref{fig:tbrowser_sub_agent}:} CUGA trajectory visualizations on API and web environments. Each illustrates the interaction between perception and action across modalities.}
\end{figure}


\section{Architecture Evolution}

\begin{figure}[htbp]
    \centering
    \includegraphics[width=1.0\textwidth]{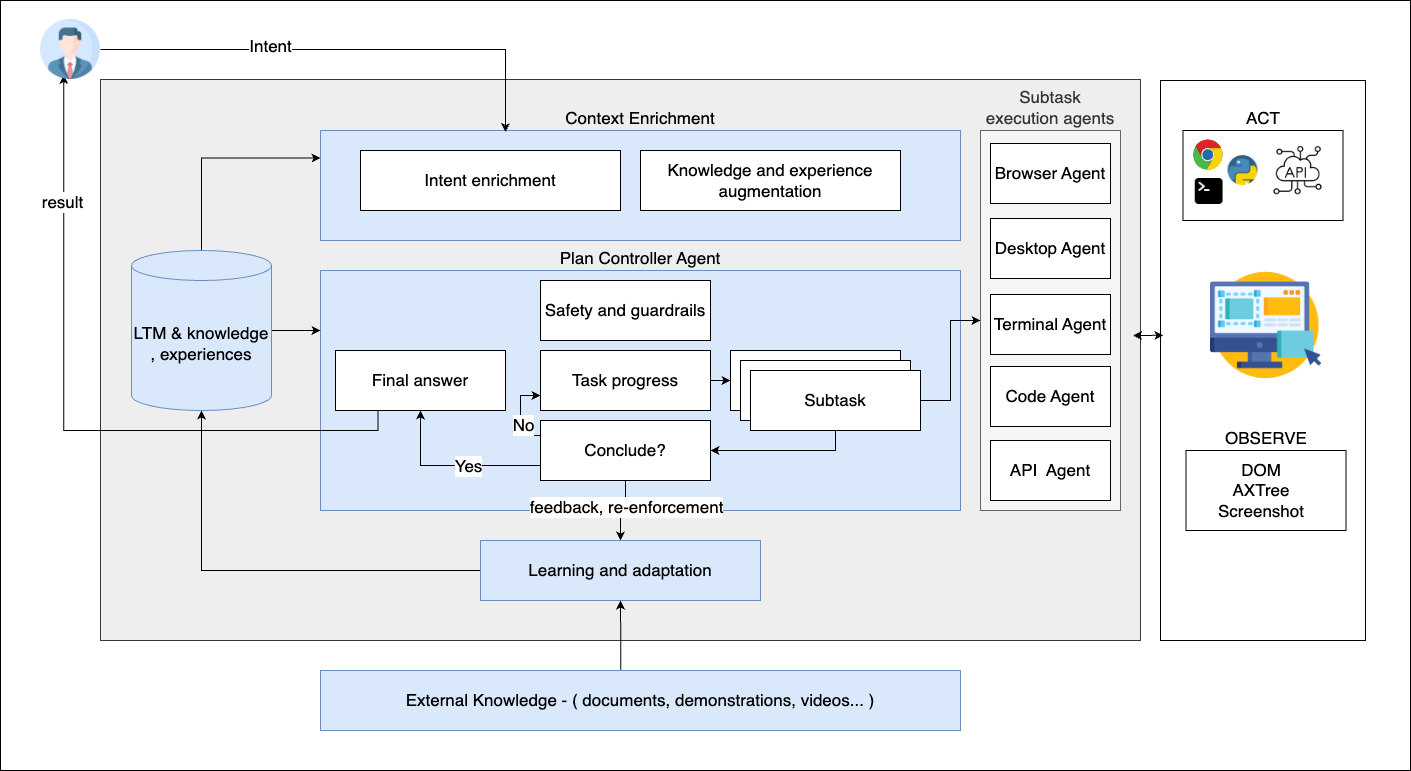}
    \caption{A simplified high-level representation of IBM CUGA architecture, illustrating the interaction between user intent, context enrichment, high-level plan controller, sub-task plan-execute agents, environment action and observation, and learning and knowledge components.} 
    \label{fig:cuga_architecture}
\end{figure}

This section details the evolutionary journey of IBM CUGA's agentic architecture. Our design philosophy centered on an iterative approach: beginning with a basic agent architecture and progressively enhancing it through rigorous failure analysis and a focus on performance-maximizing improvements. Over time, this evolved into a sophisticated multi-agent system, composed of specialized sub-agents tailored for distinct interaction modalities: web and API.
The web sub-agent is responsible for navigating and interacting with browser-based environments. It uses Playwright to control the browser and constructs its observation space from both screenshots and the accessibility tree. Evaluation against the WebArena benchmark is conducted using the official evaluation suite from BrowserGym \cite{dechezelles2024browsergymecosystemwebagent}.
The \textbf{API sub-agent} is designed to interact with structured application interfaces. It utilizes an \textbf{API Registry} to dynamically onboard and manage available applications by loading their OpenAPI schemas. This registry enables the agent to perform structured queries and hierarchical searches across APIs, scoped per application. Each application is backed by its own \textbf{MCP server}, which is automatically generated from its OpenAPI specification. This setup allows the agent to shortlist and invoke API actions in a scalable and semantically meaningful way.
The entire system is orchestrated using LangGraph\footnote{LangGraph framework - https://www.langchain.com/langgraph}, which manages stateful coordination between agents, and LangChain\footnote{LangChain - https://www.langchain.com/}, which provides a unified interface for interacting with a mix of open and frontier LLMs. A simplified high-level representation of the final architecture is depicted in Figure \ref{fig:cuga_architecture}.
In the following subsections, we provide a detailed chronicle of the key evolutionary cycles of the architecture.

\subsection{Addressing Long-Horizon and Complex Tasks}

Our initial iteration implemented a simple Plan-Act-Observe agentic loop and was evaluated on a small, representative sample of the WebArena dataset (3–5 sample templates per application domain). This baseline architecture achieved a 15\% task completion rate on this limited subset. We also evaluated the same architecture on a sample from the AppWorld benchmark training set, where it achieved only a 5\% success rate. It quickly became apparent that this approach was insufficient for handling complex, long-horizon tasks. These tasks often require the planner to orchestrate a sequence of actions and decisions while maintaining context regarding original goals and progress. Furthermore, the initial architecture struggled with tasks involving multi-site control or data flow, copy/paste operations, and the manipulation of lists and loops. It also faced significant challenges in analyzing API responses, managing variables, and shortlisting relevant APIs for task execution.

To overcome these limitations, we decomposed the planner's responsibilities across two specialized agent types:

\begin{enumerate}
    \item \textbf{Plan Controller Agent:} This agent is responsible for high-level planning, decomposing complex tasks into sub-tasks, selecting optimal sub-task sequencing, handling loops and lists, tracking sub-task progress, determining task completion, and passing variables between sub-tasks to maintain continuity and context across the execution flow.
    \item \textbf{Sub-task Plan-Execute Agents:} These agents focus on the local planning of individual steps, UI element grounding (locating), and interaction. There are two types of sub-task agents: \textit{browser-based} agents, which interact with web interfaces, and \textit{API-based} agents, which handle API calls, response parsing, and variable management.
\end{enumerate}

This decomposition enabled more robust handling of complex tasks, as the Plan Controller Agent managed the overall task flow. At the same time, the Sub-task Plan-Execute Agents handled the specifics of interacting with either the user interface (UI) or APIs. This separation of concerns significantly improved performance and laid the groundwork for further architectural refinements.  

\subsection{Sub-Execution Agents Challenges and Enhancements}

\begin{figure}[htbp]
    \centering
    \begin{subfigure}[t]{0.48\textwidth} 
        \centering
        \includegraphics[width=\textwidth]{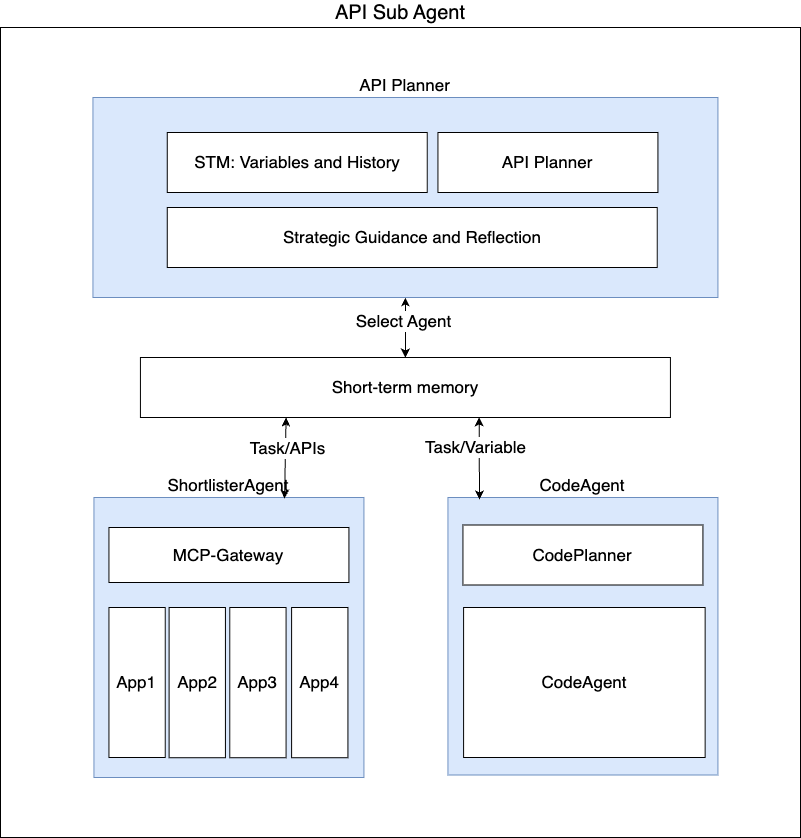}
        \caption{This diagram outlines the architecture of an API Sub Agent, featuring an API Planner that coordinates task execution through short-term memory, strategic guidance, and variable management. It connects to two agents: the ShortlisterAgent, which selects applications and APIs via an MCP-Gateway, and the CodeAgent, which manages code tasks through a nested CodePlanner and CodeAgent that returns variables, enabling dynamic, context-aware decision-making across the system.}
        \label{fig:api_sub_agent}
    \end{subfigure}
    \hfill 
    \begin{subfigure}[t]{0.48\textwidth} 
        \centering
        \includegraphics[width=\textwidth]{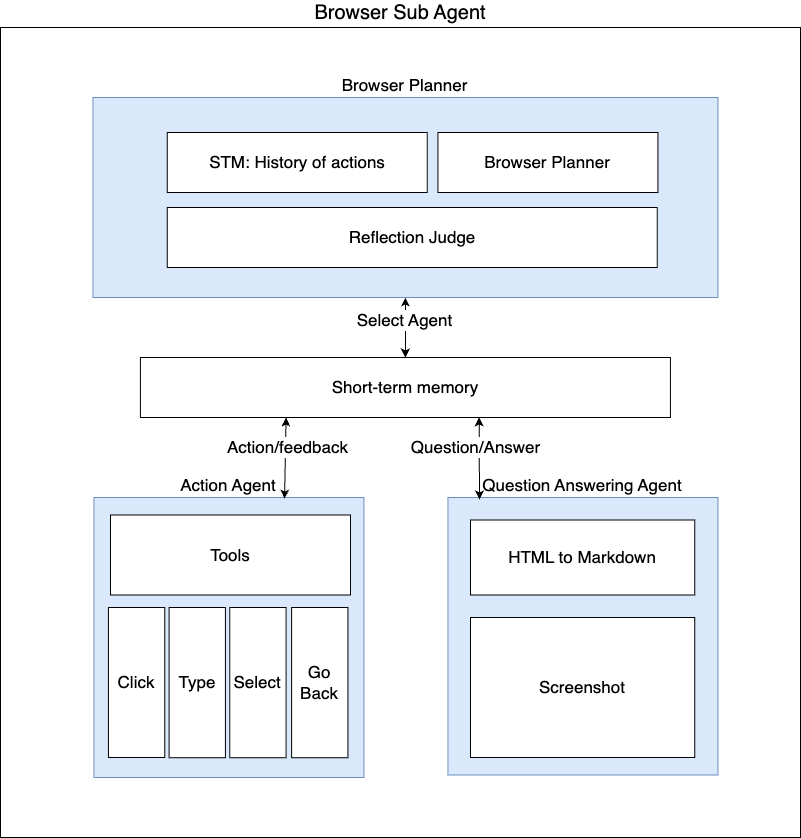}
        \caption{This diagram presents the architecture of a Browser Sub Agent, structured around a central Browser Planner that includes a short-term memory (STM) for action history, a planning module, and a Reflection Judge for evaluating decisions. The system connects to a Select Agent and two specialized agents: the Action Agent, which executes browser interactions using tools such as Click, Type, Select, and Go Back; and the Question Answering Agent, which answers questions about the current page by processing web content through HTML-to-Markdown conversion and capturing screenshots. The architecture enables dynamic, memory-informed browsing and content extraction.}
        \label{fig:browser_sub_agent}
    \end{subfigure}
    \caption{Architecture of the API and Browser Sub Agents}
    \label{fig:two_images}
\end{figure}

Building upon our prior iteration, we refined our sampling methodology to construct a more comprehensive and representative subset of the WebArena benchmark. This expanded dataset, comprising 44 data points spanning both previously encountered and novel task types, enabled a more rigorous evaluation. On this refined benchmark, our system achieved a task completion rate of 37.8\%. However, detailed failure analysis revealed persistent limitations across both \textit{Browser} and \textit{API} Sub-task Plan-Execute agents.

\subsubsection{Browser Sub-task Agent Challenges}

\begin{enumerate}
    \item \textbf{Action Execution Discrepancies:} While the planner frequently identified the correct high-level action (e.g., selecting from a dropdown, entering text), the Action agent often failed to translate this into precise interaction steps due to the heterogeneity and non-standard implementations of UI components across websites.
    \item \textbf{UI Element Grounding Failures:} The agent frequently misidentified or failed to locate the intended UI element, resulting in execution errors.
    \item \textbf{Complex Interaction and Extraction:} Combining UI interaction and complex information extraction within a single prompt proved overly burdensome, necessitating a decoupling of perception and action responsibilities.
    \item \textbf{Popup Obstruction:} Pop-up elements occasionally occluded the observation space, impairing the agent's ability to perceive and interact with the environment effectively.
\end{enumerate}

\subsubsection{Enhancements for Browser Sub-task Agent}

\begin{enumerate}
    \item \textbf{Robust Element Interaction:} We introduced a feedback loop within the Action agent to iteratively refine interaction strategies, enabling it to bypass obstructive or irrelevant UI elements (e.g., popups) and adapt to diverse UI implementations.
    \item \textbf{Dedicated Information Extraction Agent:} A specialized agent was introduced to handle information extraction, operating on a distinct observation space. This separation improved both interaction fidelity and extraction accuracy.
    \item \textbf{Enhanced Visual Context:} The agent's perceptual input was augmented with screenshots alongside the accessibility tree, providing richer visual grounding for decision-making and UI element localization.
\end{enumerate}

\subsubsection{API Sub-task Agent Challenges}

\begin{enumerate}
    \item \textbf{API Shortlisting Failures:} The agent frequently failed to identify the most relevant API for a given sub-task, particularly in the presence of verbose or ambiguous OpenAPI specifications.
    \item \textbf{Variable Management Deficiencies:} The agent lacked robust mechanisms for tracking and propagating variables across API calls and sub-tasks, leading to execution inconsistencies.
    \item \textbf{Unexpected Output Handling:} The agent struggled to interpret and respond to unanticipated outputs from API calls or code execution, such as filtering or transforming intermediate results.
    \item \textbf{Lack of Global Planning Context:} The agent operated in a myopic, step-wise fashion, lacking the ability to reason about global task structure or interdependent API workflows.
\end{enumerate}

\subsubsection{Enhancements for API Sub-task Agent}

\begin{enumerate}
    \item \textbf{API Registry and Shortlisting:} We developed an API registry that maps applications to their available APIs, enabling the agent to shortlist relevant APIs based on the sub-task and application context.
    \item \textbf{Concise OpenAPI Representation:} To reduce token overhead and improve interpretability, OpenAPI specifications were transformed into a minimized, LLM-friendly format that preserved essential semantics while omitting extraneous metadata.
    \item \textbf{Variable Propagation Mechanism:} A system-wide variable tracking framework was introduced, allowing the Plan Controller Agent to pass and update variables across sub-tasks, ensuring coherent multi-step execution.
    \item \textbf{Reflective Execution Loop:} The API agent was augmented with a reflection mechanism, enabling it to re-evaluate and revise its strategy in response to unexpected outputs or execution failures.
\end{enumerate}

\subsection{Enhancing Stability and Mitigating Hallucinations}

Large Language Models and agentic AI systems inherently exhibit variability in their execution. Running the same code on the same benchmark and environment can yield diverse reasoning paths and action sequences. While this characteristic fosters creative problem-solving, alternative exploration, and diverse decision-making, it can also lead to inconsistencies and hallucinations. To mitigate these issues and enhance stability, we incorporated reflection, critique, and judgment techniques within several of our agents. These techniques were implemented through both prompt engineering, which involved refining existing prompts to encourage self-assessment, and the introduction of dedicated reflection and judgment agents. This dual approach allowed us to address both the underlying reasoning processes and the final outputs, improving the reliability and consistency of CUGA's performance. 

\subsection{Planner Alignment through Context Enrichment, Learning, and Knowledge Injection}

Further failure analysis revealed that both the high-level Plan Controller and the Sub-task Planners occasionally struggled due to a lack of relevant application knowledge. This knowledge deficit led to misinterpretations of sometimes vague user intents, resulting in incorrect or ineffective planning and an inability to recover from flawed initial reasoning or planning decisions. To address these challenges, we introduced a context curation and knowledge injection layer responsible for the following:

\begin{enumerate}
    \item \textbf{User Utterance Processing:} This component assesses the quality of user utterances and paraphrases unclear or ambiguous requests. This ensures that the planners receive well-defined and actionable intents.
    \item \textbf{Application Navigation Knowledge Acquisition (for browser sub-agent):} For each newly encountered application, this component explores the application's navigation space, effectively mining a site-map-like structure. This structural knowledge enriches the context for intents that require a comprehensive understanding of the application's functionality, guiding the planners toward appropriate actions.
    \item \textbf{Contextual Enrichment based on Task Assessment (for browser sub-agent):} Based on the assessed task, this component injects relevant application navigation knowledge and other contextual information, further refining the planners' understanding of the task requirements and the available tools.
\end{enumerate}

This knowledge injection and context enrichment layer significantly improved the alignment of the planners with user intents and application functionality. By providing a richer understanding of the application landscape and clarifying user requests, the planners were able to make better and more informed decisions, leading to more effective and robust task execution.

\section{Results}

Figure \ref{fig:cuga_evolution} illustrates the overall performance improvement achieved through the iterative evolution of our architecture. Each iteration involved two evaluations: first, a validation on the previously used sample to confirm the impact of enhancements aimed at addressing observed failures, and second, a test on a larger, more representative sample to assess generalizability. The description of the samples we used is detailed in Table \ref{table:samples}.  It is worth noting that, in some instances, increasing the sample size has resulted in a slight reduction in performance. This phenomenon reflects the inherent approximation of our sampling strategy, where smaller samples may not fully capture the complexity of the benchmark. Despite these minor fluctuations, the graph demonstrates a clear and consistent trend of improvement in task completion rate over time. This upward trajectory highlights the effectiveness of our iterative refinement process, emphasizing the value of continuous analysis and targeted enhancements in developing a robust and high-performing agentic system.

\begin{table}[h!]
\centering
\caption{Progressively larger WebArena samples used in iterative development, enabling failure-driven refinement and generalization assessment across increasing benchmark coverage.}
\begin{tabular}{lcc}
\toprule
\textbf{Sample Name} & \textbf{Sample Size} & \textbf{Description} \\
\midrule
Initial & 22 & Initial representative templates per domain \\
Nano & 44 & Improved larger sample distribution of success and failure \\
Micro & 90 & 50\% coverage for templates \\
Mini & 190 & All templates \\
Full & 812 & Full benchmarks \\
\bottomrule
\end{tabular}
\label{table:samples}
\end{table}

\begin{figure}[H]
    \centering
    \begin{tikzpicture}
        \begin{axis}[
            xlabel=Evaluation Run,
            ylabel=Task Completion Rate (\%),
            xmin=0, xmax=9,
            ymin=0, ymax=70,
            xtick={1,2,3,4,5,6,7,8,9},
            xticklabels={1-initial, 1-nano, 2-nano, 2-micro, 3-micro, 3-mini, 4-mini, 4-full, 5-full},
            xticklabel style={rotate=45,anchor=east},
            grid=major,
            legend pos=north west,
            legend style={font=\footnotesize},
        ]
        \addplot[color=blue, mark=*] table [x index=0, y index=1] {
            0 15.2
            1 26.0
            2 37.8
            3 43.2
            4 45.5
            5 42.4
            6 55.5
            7 50.1
            8 61.7
            
        };
        \addlegendentry{Task Completion Rate}

        \addplot[only marks, mark=*, color=red, scatter,
            scatter src=explicit symbolic,
            scatter/classes={
                22={mark size=2pt},
                44={mark size=4pt},
                90={mark size=5pt},
                190={mark size=7pt},
                812={mark size=9pt}
            }] table [x index=0, y index=1, meta index=2] {
            0 15.2 22
            1 26.0 44
            2 37.8 44
            3 43.2 90
            4 45.5 90
            5 42.4 190
            6 55.5 190
            7 50.1 812
            8 61.7 812
        };
        \addlegendentry{Sample Size}
        \end{axis}
    \end{tikzpicture}
    \caption{Evolution of CUGA Architecture Performance. The graph illustrates the task completion rate (\%) across different evaluation runs. Each run is denoted by the iteration and the dataset sample name. The size of the markers corresponds to the sample size, where in each iteration we validate performance gains on the previous sample, as well as enlarge the sample to learn about new failures. Note that in some instances, increasing the sample size led to a minor reduction in performance, reflecting the approximation inherent in our sampling strategy. Overall, the graph demonstrates a clear trend of improvement in task completion rate over time, showcasing the effectiveness of our iterative refinement process.}
    \label{fig:cuga_evolution}
\end{figure}

The AppWorld benchmark provides a structured and multi-layered evaluation of agents' ability to interact with API-driven applications across increasing levels of complexity. As shown in Table~\ref{tab:appworld_results}, CUGA achieves a task goal completion rate of 73.2\% under normal conditions and 57.6\% in the more challenging settings. At the scenario level — which requires successful completion of all task instances within the same task template, usually 3 different instances — the agent reaches 62.5\% and 48.2\% respectively, reflecting strong generalization across diverse real-world tasks. Notably, CUGA is now the state-of-the-art performer on the AppWorld benchmark, ranking first on the official leaderboard.

Performance across levels reveals important trends. On Level 1 scenarios, which typically involve single app tasks, 2-3 API calls or linear workflows, the agent performs nearly flawlessly, achieving over 91\% task goal completion and 84–87\% scenario completion. This indicates the robustness of CUGA's sub-agent decomposition and basic API reasoning capabilities. As complexity increases in Levels 2 and 3 - where tasks require variable tracking, conditional logic, and interaction across multiple APIs—performance degrades more gradually. The agent still maintains above 68\% scenario completion on Level 2 (normal), but dips to 38.1–38.5\% at Level 3, revealing key areas for further improvement in reasoning over branching workflows and handling unexpected outputs.

The average number of interactions per scenario also increases with difficulty, suggesting that CUGA adapts its strategy to the problem space, rather than simply following static plans. This is particularly evident in the challenge split, where the agent demonstrates a more efficient interaction pattern in early levels (fewer steps), but needs longer trajectories to resolve ambiguities or recover from partial failures in more complex scenarios. Overall, these results demonstrate that CUGA not only achieves strong performance in controlled web environments (WebArena), but also generalizes effectively to dynamic API-based tasks (AppWorld). 

\begin{table}[ht!]
    \centering
    \caption{CUGA performance on the AppWorld benchmark across increasing levels of API complexity, showing strong task and scenario completion rates and adaptive interaction strategies under both normal and challenge settings.}
    \label{tab:appworld_results}
    \vspace{0.5em}      
    \begin{tabular}{l cc cc cc}
        \toprule
        \multirow{2}{*}{\textbf{Level}} &
        \multicolumn{2}{c}{\textbf{Task Goal Completion (\%)}} &
        \multicolumn{2}{c}{\textbf{Scenario Goal Completion (\%)}} &
        \multicolumn{2}{c}{\textbf{Avg.\ Interactions}} \\[0.15em]
        \cmidrule(lr){2-3}\cmidrule(lr){4-5}\cmidrule(lr){6-7}
        & Normal & Challenge & Normal & Challenge & Normal & Challenge \\
        \midrule
        All     & 73.2 & 57.6 & 62.5 & 48.2 & 10.69 & 8.40 \\
        Level 1 & 91.2 & 91.7 & 84.2 & 87.5 &  5.94 & 4.65 \\
        Level 2 & 77.1 & 58.7 & 68.8 & 42.0 & 10.36 & 8.33 \\
        Level 3 & 54.0 & 44.1 & 38.1 & 38.5 & 12.69 & 11.86 \\
        \bottomrule
    \end{tabular}
\end{table}

\section{Lessons Learned, Challenges, and Roadmap}

\paragraph{Methodology and Tools:} Our iterative development methodology, coupled with smart evaluation and analytics tools, has been crucial for rapid progress and achieving state-of-the-art results cost-effectively. These tools have enabled us to assess failures and identify areas for improvement effectively. However, the process of finding, classifying, and aggregating root causes of failures within agentic trajectories remains labor-intensive, even with current tooling. Automating some of this manual effort by leveraging AI agents could significantly accelerate the development of agentic systems and potentially pave the way for fully autonomous evolution. Another interesting research opportunity lies in the area of smart regressions and evaluating individual agents within a comprehensive agentic system. Currently, best practices and lessons learned in this area are not widely shared. We hope this paper contributes to the community by sharing our experiences. Within IBM, we are leveraging insights, methodologies, and tools from initiatives like IBM CUGA and others, integrating and hardening them into the observability and analytics layers of the IBM Agentic Middleware platform. This will allow clients to benefit from these capabilities at a product-grade level.

\paragraph{Evaluating Generalist Agents on Realistic Benchmarks:} Creating and maintaining effective benchmarks is a significant challenge. Many existing benchmarks are developed within academic settings, reflecting the tasks and tools readily available to researchers. While valuable, these benchmarks often lack the complexity and nuances of real-world scenarios. A promising recent effort, TheAgentCompany \cite{xu2024theagentcompanybenchmarkingllmagents}, aims to address this by simulating more realistic tasks and incorporating human-agent communication. However, current benchmarks often focus on "happy path" scenarios and overlook critical aspects, such as safety, particularly for agents interacting with enterprise systems, applications, and APIs. They typically do not evaluate agent behavior in exceptional circumstances or when human-in-the-loop interaction is required. A notable effort in this direction is STWebAgentBench \cite{levy2024st}, which extends WebArena by incorporating safety and policy adherence measurements alongside task completion. Ultimately, realistic benchmarks should mirror the specific environments, policies, and enterprise use cases of individual organizations.

At IBM Research, we plan to further evolve CUGA to be a safe and policy-compliant, as well as a collaborative and trustworthy AI agent with human-in-the-loop support. Furthermore, we plan to evaluate CUGA on several leading benchmarks. These benchmarks combine multiple tools, desktop applications, coding, and APIs. We will also contribute to the community's efforts in developing more comprehensive evaluation frameworks. In particular, given the structure and scope of the AppWorld benchmark, we plan to explore the dimension of \textit{API safety}, including how agents can reason about, restrict, and validate API usage in alignment with enterprise policies and security constraints. The AppWorld benchmark stands out as a remarkably well-designed and comprehensive evaluation suite, thoughtfully constructed to cover a wide range of realistic use cases across multiple applications. Its inclusion of diverse task types and multi-application workflows makes it a valuable resource for advancing the development of robust and enterprise-ready agentic systems.

\paragraph{Learning, Customization, and Adaptation:}  Traditional system development relies on programming languages and low-code/no-code tools to build and test systems according to specifications defined in requirements documents. Agentic systems offer a radically different paradigm. Agents can learn to perform tasks much like humans: by studying documentation and policies, receiving instructions, observing expert behavior, analyzing videos, and through interactive discovery and safe exploration. Emerging research suggests this vision is within reach. At IBM Research, we have begun experimenting with methods to enable CUGA to learn from unstructured documents, empowering non-technical users to configure and customize its behavior to meet their specific needs.

\paragraph{Smaller and Open Models:} While our current state-of-the-art results are based on a frontier model, this choice was primarily driven by time efficiency, allowing us to iterate on our agentic architecture across various benchmarks rapidly. We recognize the significant advantages of smaller and open models, particularly in terms of accessibility, affordability, privacy, efficiency, and cost. We have initiated experiments to evaluate our architecture using these models and plan to publish a comprehensive evaluation of our findings upon completion.

\section{Conclusion}

In this paper, we have presented the current status of our work on developing IBM CUGA. We detailed the iterative evolution of our agentic architecture, the methodology and tools that facilitated rapid learning and cost-effective improvements, and key lessons learned and challenges ahead. Our results demonstrate the effectiveness of our approach, achieving state-of-the-art performance on both the WebArena and AppWorld benchmarks. Our contributions offer valuable insights and practical guidance to the community, paving the way for future advancements in the field of enterprise-ready, agentic AI systems.

\bibliographystyle{plain}
\bibliography{references} 

\end{document}